\documentclass[aps,pre,reprint,groupedaddress,showpacs]{revtex4-1}
\usepackage{graphicx}
\usepackage{amssymb}
\usepackage[normalem]{ulem}
\usepackage{xcolor}

\begin{document}

\title{Jammed Spheres: Minkowski Tensors Reveal Onset of Local Crystallinity}

\author{Sebastian C.~Kapfer$^{1}$}
\email{Sebastian.Kapfer@physik.uni-erlangen.de}
\author{Walter~Mickel$^{1,2}$}
\email{Walter.Mickel@physik.uni-erlangen.de}
\thanks{SCK and WM have contributed equally to this work.}
\author{Klaus~Mecke$^1$}
\email{Klaus.Mecke@physik.uni-erlangen.de}
\author{Gerd~E.~Schr\"oder-Turk$^1$}
\email{Gerd.Schroeder-Turk@physik.uni-erlangen.de}
\affiliation{$^1$ Institut f\"ur Theoretische Physik, FAU Erlangen-N\"urnberg, Staudtstr.~7, D-91058 Erlangen, Germany}
\affiliation{$^2$ Universit\'{e} de Lyon, F-69000, Lyon, France; CNRS, UMR5586, Laboratoire PMCN, Lyon, France;\\
Karlsruhe Institute of Technology, Institut f\"ur Stochastik, Kaiserstr.~89, D-76128 Karlsruhe, Germany}

\date{\today}

\begin{abstract}
The local structure of disordered jammed packings of monodisperse spheres without friction,
generated by the Lubachevsky-Stillinger algorithm, is studied for packing fractions above
and below $64\%$. The structural similarity of the particle environments to fcc
or hcp crystalline packings ({\em local crystallinity}) is quantified by order
metrics based on rank-four Minkowski tensors. We find a critical packing
fraction $\phi_{\rm c}\approx 0.649$, distinctly higher than previously reported
values for the contested random close packing limit. At $\phi_{\rm c}$, the
probability of finding local crystalline configurations first becomes finite
and, for larger packing fractions, increases by several orders of magnitude.
This provides quantitative evidence of an abrupt onset of local crystallinity
at $\phi_{\rm c}$.
We demonstrate that the identification of local crystallinity by
the frequently used local bond-orientational order metric $q_6$ produces
false positives, and thus conceals the abrupt onset of local crystallinity.
Since the critical packing fraction is significantly above results from mean-field analysis of the mechanical contacts for frictionless spheres, it is suggested that
dynamic arrest due to isostaticity and the alleged geometric phase
transition in the Edwards framework may be disconnected phenomena.
\end{abstract}


\maketitle 

Now classic experiments showed that disordered sphere packings can only be prepared up to a maximum packing fraction of $\phi_{\rm RCP}\approx 0.64$ \cite{BritJApplPhys69_Scott,Bernal:1959}. 
This packing fraction, referred to as \emph{random close packing} (RCP), is substantially lower than the packing fraction $\phi_{\rm fcc}\approx 0.74$ of the densest crystalline sphere packing. Maximal packing fractions close to $\phi_{\rm RCP}$ have been shown for several experimental protocols \cite{PRE05_Aste}; protocols inducing local crystallization are able to reach higher packing fractions \cite{EPJE00_Nicolas}.
Numerical protocols to generate static sphere packings both below and above 0.64 are the Lubachevsky-Stillinger (LS) algorithm \cite{SkogeDonevStillingerTorquato:2006} and the Jodrey-Tory algorithm \cite{PRA85_Jodrey}. 

The nature and the existence of a transition near $\phi_{\rm RCP}$ are disputed. As sphere configurations with packing fractions between $\phi_{\rm RCP}$ and $\phi_{\rm fcc}$ evidently exist \cite{PRE02_Kansal}, an alleged structural transition must be due to either a vanishing configuration space density of {these states} or to the inability to reach these within the considered ensemble or by the given dynamics. 
Within the framework of equilibrium (thermal) hard spheres,
the concept of RCP has been related to the terminus of the branch of metastable states avoiding crystallization; divergence of pressure is reported to occur at $\phi=0.640\pm 0.006$ \cite{KamienLiu:2007} and $\phi\approx 0.65$ \cite{EPL04_Aste}.  
By contrast, in an athermal statistical ensemble where the role of energy is played by volume \cite{PhysicaA89_Edwards}, a mean-field study based on mechanical contact numbers has reported $\phi_{\rm RCP}\approx 0.634$ \cite{nature08_Song}. 
An order/disorder transition in an athermal ensemble
has been demonstrated for a lattice model \cite{Radin:2008,*AristoffRadin:2010}.
Support for the phase transition scenario is deduced from the fact that
the volume fraction of polytetrahedra increases with packing
fraction up to $\phi\approx 0.646$ and then decreases, as these
structures transform into crystalline order
\cite{PRL07_Anikeenko,AnikeenkoMedvedevAste:2008}.

In addition to these phase transition scenarios, where $\phi_{\rm RCP}$ is
interpreted as the density of the disordered phase at coexistence,
the notion of the maximally random jammed state (MRJ, \cite{PRL00_Torquato}) has been proposed (as the maximally disordered state amongst all jammed packings); with respect to a number of common measures of order, the MRJ packing fraction has been estimated as $\phi_{\rm MRJ}\approx 0.63$ \cite{PRE02_Kansal}. 

The resolution of the RCP problem relies on suitably defined order metrics to quantify packing structure. A common approach to local structure characterization is by analysis of nearest neighborhoods \cite{PRB83_Steinhardt,Bargiel:2001:0921-8831:533}. Often, the bond-orientational order metrics $q_l$ defined by Steinhardt {\em et al.~}\cite{PRB83_Steinhardt} are applied to sphere packings  \cite{PRE02_Kansal,EPJE10_Xu}. However, these and other neighborhood-based order metrics {\cite{FakenJonsson:1994,*AcklandJones:2006}}
have shortcomings. First, they suffer from the ambiguous definition of the nearest neighborhood
\cite{MickelKapferSchroeder-Turk:2011}. Second, in their common use as single scalar order metrics \cite{PRE02_Kansal,EPJE10_Xu}, {they are insufficient} to conclusively distinguish order and disorder \cite{KlumovKhrapakMorfill:2011,*KlumovUspekhi:2010}. A non-negligible fraction of non-crystalline {environments} is often {incorrectly} identified as crystalline ({\em false positives}), since their $q_6$ are close or identical to the {reference $q_6$ values in crystals}.

\begin{figure}[t]
\begin{center}
\begin{tabular}{cc}
 \includegraphics[bb=470 187 1240 820, clip, width=0.49\columnwidth]{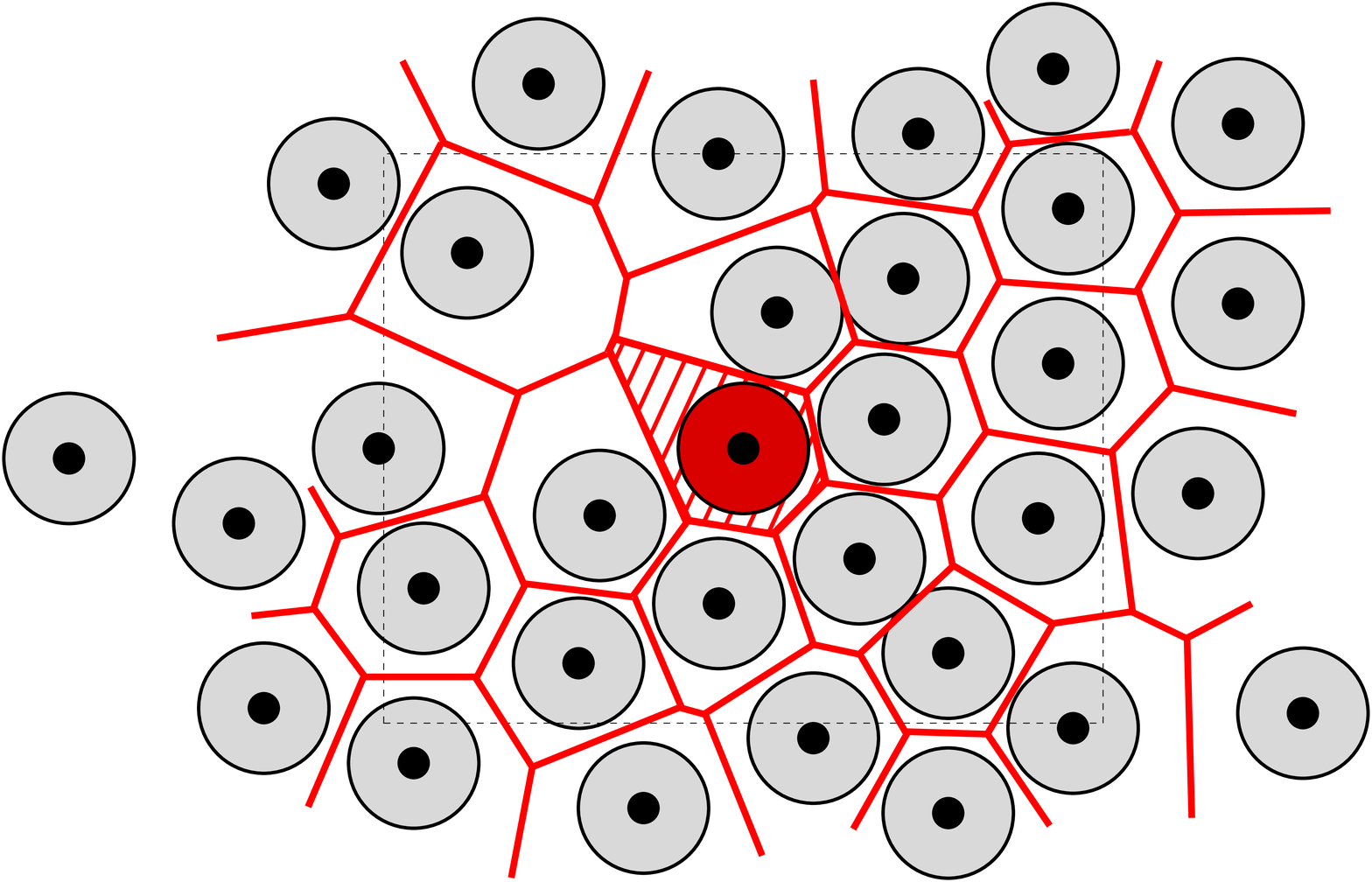} & 
 \includegraphics[bb=470 187 1240 820, clip, width=0.49\columnwidth]{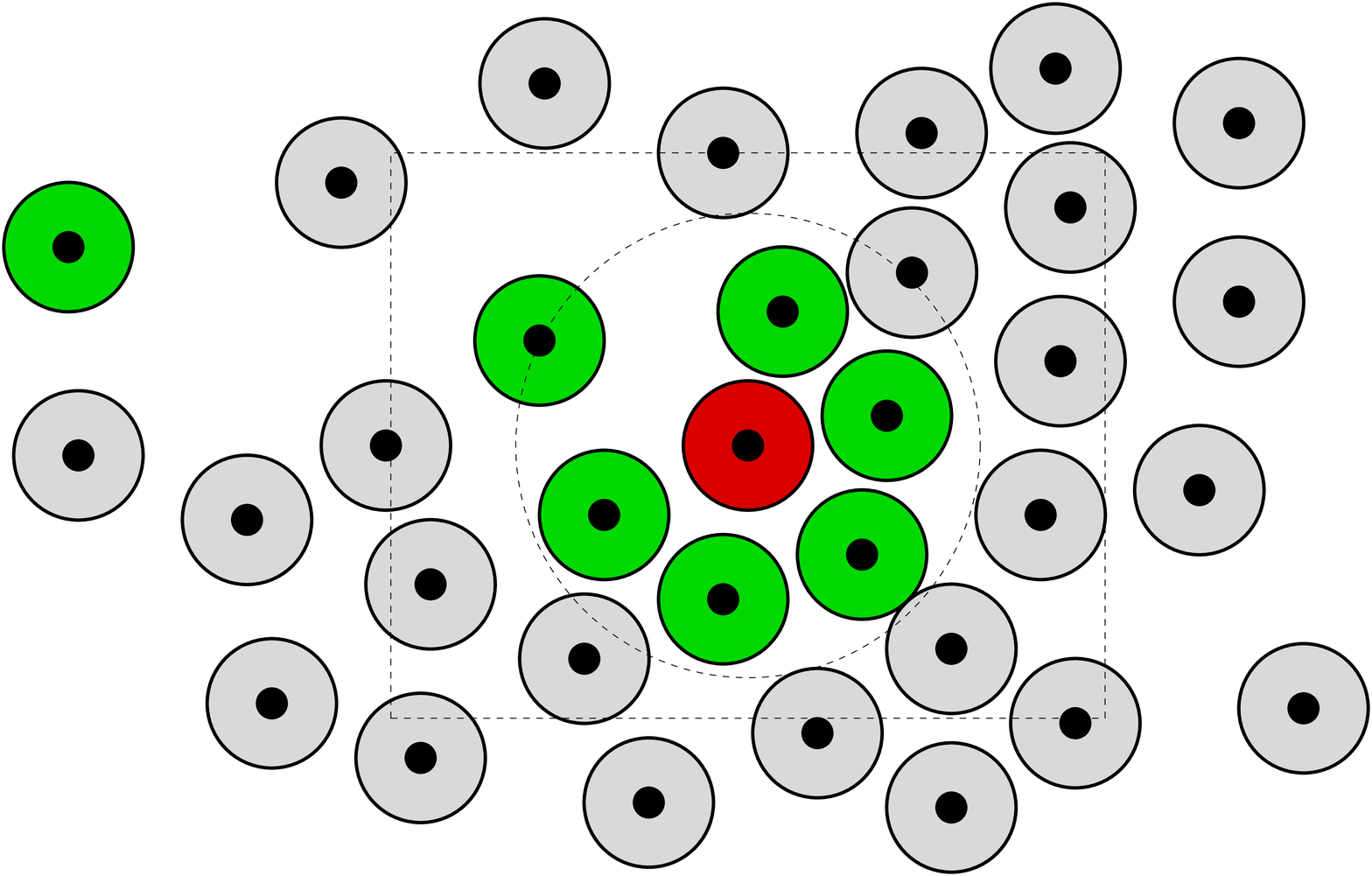} \\
\end{tabular}
\end{center}
\caption{
\label{fig:neighbors-voronoi} (Color online) (left) The Voronoi diagram of a packing associates convex cells with the individual particles. Each Voronoi cell $K$ contains one particle $k$ such that the distance of any point $p\in K$ to particle $k$ is smaller than the distance of $p$ to any other particle. (right) For the evaluation of $q_6$, {we define} the set of nearest neighbors of particle $k$ as {those} 12 other particles (6 in 2D) which are closest to $k$. }
\end{figure}

Alternatively, the structure of monodisperse sphere packings can be characterized by analysis of the Voronoi {cells} of the particle centers, see Fig.~\ref{fig:neighbors-voronoi}. Suitable morphological descriptors, such as Minkowski tensors \cite{EPL10_SchroederTurk,KapferFluids:2010,*AdvMat11_SchroederTurk}, can then be used to quantify the cell shape and hence the local structure.
Here, we show that {\em crystalline order metrics} can be constructed from rank-four Minkowski tensors of the Voronoi cells
that give stringent criteria for fcc or hcp crystalline order. 
For jammed sphere packings generated by the LS algorithm,
these order metrics reveal an abrupt onset of crystallinity at a critical packing fraction $\phi_{\rm c}\approx 0.649$.

Eigenvalue ratios of rank-two Minkowski tensors quantify anisotropy of the
particle environments in jammed bead packs \cite{EPL10_SchroederTurk}. The Voronoi
cells of the fcc and hcp close packing are isotropic, in the terminology of
Ref.~\cite{EPL10_SchroederTurk}, while cells found in disordered packings typically are
not.  However, rank-two tensors are insufficient to distinguish different types of
isotropic cells.
These cell types can be classified using the rank-four Minkowski tensor $W_1^{0,4}$.
The tensor $W_1^{0,4}$ of a Voronoi cell is given as
the sum of tensor products of the facet normals, weighted by the facet areas,
\begin{equation}
 (W_1^{0,4})_{ijkl}:= \frac 1A\cdot \sum_{f} a(f)\, n_i n_j n_k n_l,
 \label{eq:def-w104}
\end{equation}
where $n_i:=(\mathbf{n}(f))_i$ with $i=1,2,3$ are the cartesian components of the facet normal, and
$a(f)$ is the surface area of facet $f$; further, $A:=\sum_{f}a(f)$ is the total surface area;
all sums run over the facets of the Voronoi cell $K$.
In close analogy to the stiffness tensor of continuum mechanics, symmetry under permutations of indices allows the reduction of $W_1^{0,4}$ to a $6\times 6$ symmetric matrix \footnote{In contrast to the elastic tensor, the Minkowski tensor obeys also the relation $(W_1^{0,4})_{ijkl}=(W_1^{0,4})_{ikjl}$.}. The six eigenvalues $(\varsigma_1,\ldots,\varsigma_6)$ of this matrix are dimensionless due to normalization by $A^{-1}$ and are rotational invariants \cite{MehrabadiCowin}. A concise quantitative measure for the similarity of a given Voronoi cell $K$ to the Voronoi cell $K_{\mathrm{fcc}}$ of a crystalline {fcc packing} is given by the {\em fcc crystalline order metric}
\begin{equation}\label{eq:fingerprint_fcc}
 \Delta_\mathrm{fcc}(K):=\left[\sum_{i=1}^6 \Bigl(\varsigma_i(K)-\varsigma_i(K_{\mathrm{fcc}})\Bigr)^2 \right]^{1/2}.
\end{equation}
An analogous order metric $\Delta_\mathrm{hcp}$ is defined for hcp cells.

\begin{figure}[t]
  \centering
\includegraphics[bb=50 51 266 185, clip, width=\columnwidth]{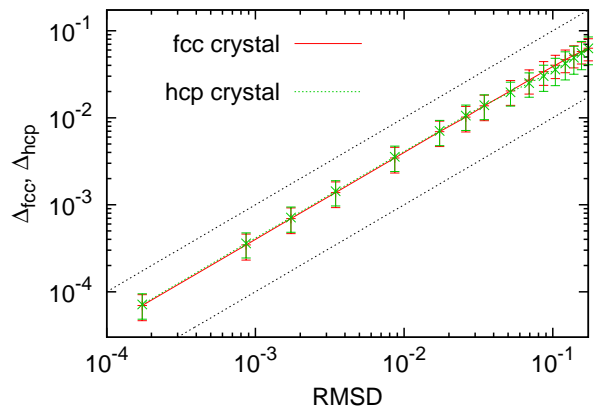}
\caption{\label{fig:einstein-solid-dependence}(Color online) The relationship between root mean square particle displacement
from an ideal fcc (red solid line) and hcp (green dashed) lattice and 
and the corresponding values of $\Delta_{\mathrm{fcc}}$ and $\Delta_{\mathrm{hcp}}$ respectively
is approximately linear. Error bars represent standard deviations among cells due to the statistical
displacement of the particles. The dashed curves above and below the data are linear functions with slope
1 and 0.1 respectively.}
\end{figure}

Figure~\ref{fig:einstein-solid-dependence} supports our claim that $\Delta_{\mathrm{fcc}}$ and $\Delta_{\mathrm{hcp}}$ measure deviations of a Voronoi cell's shape from the ideal fcc or hcp cell. The vertices of an ideal fcc or hcp lattice are displaced by small random vectors. Figure~\ref{fig:einstein-solid-dependence} shows averages and standard deviations (as error bars) of $\Delta_{\mathrm{fcc}}$ and $\Delta_{\mathrm{hcp}}$ as function of the root mean square displacement (RMSD) from the ideal lattice points, demonstrating an approximately linear relationship between deviations from the ideal crystalline shapes and the crystalline order metrics $\Delta_{\mathrm{fcc}}$ and $\Delta_{\mathrm{hcp}}$. The quantitative agreement between the functions for hcp and fcc cells justifies the use of the same threshold value {for selecting both fcc- and hcp-like cells from a packing.}

\begin{figure}[t]
\centering
\includegraphics[width=\columnwidth]{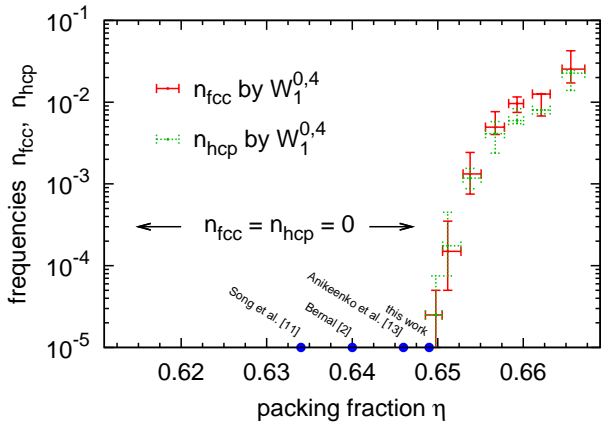}
\caption{\label{fig:frequ-fcc-hcp-fct-phi}
(Color online) Fractions $n_{\mathrm{fcc}}(\phi)$ (red) and $n_{\mathrm{hcp}}(\phi)$ (green) of fcc- and hcp-like cells as function of the packing fraction $\phi$ (cell types identified by $W_1^{0,4}$).
Below $\phi_{\rm c}\approx 0.649$, crystallinity is negligeable, and was found, amongst 3000 simulations with $\phi\in [0.56,\phi_{\rm c}]$, with rates {substantially} smaller than the inverse system size.
Above $\phi_{\rm c}$, the probability of crystalline cells increases by orders of magnitude.
Each data point is computed from $M\approx 20$ packings of $4\times 10^4$ spheres each.
Horizontal error bars correspond to the variations in $\phi$ encountered for the same growth rate $\gamma$ of the LS algorithm \cite{SkogeDonevStillingerTorquato:2006}; vertical error bars represent the interval {$([ N_{\mathrm{fcc}}]_{0.25}, [ N_{\mathrm{fcc}}]_{0.75})$}.
}
\end{figure}

The crystalline order metrics $\Delta_{\rm fcc}$, $\Delta_{\rm hcp}$ are used to identify crystalline clusters
in jammed sphere packings generated by the Lubachevsky-Stillinger protocol
\cite{SkogeDonevStillingerTorquato:2006}.
Figure \ref{fig:frequ-fcc-hcp-fct-phi} shows, as key result of this paper, that
(a) crystalline fcc and hcp order is absent for packing fractions below a critical value
$\phi_{\rm c}\approx0.649$; and that (b) above $\phi_{\rm c}$ the fraction of
crystalline fcc or hcp in LS simulations is non-zero and rapidly increases
by several orders of magnitude.
As expected for an athermal system, no systematic difference between the number
of hcp and fcc cells is observed, in contrast to crystallization dynamics in
equilibrium \cite{nature1997_bolhuis,*nature1997_woodcock}.

We measure the fraction of fcc cells as
$n_{\mathrm{fcc}}(\phi)=[ N_{\mathrm{fcc}}]_{0.5}/N$, where $N_{\mathrm{fcc}}$ is the number
of cells with $\Delta_{\rm fcc}<\delta=0.005$, in a sample of $N= 4\times 10^4$ spheres.
$[ N_{\mathrm{fcc}}]_{0.5}$ {is the median of $N_{\rm fcc}$} over $M\approx 20$ {packings of similar $\phi$, see Fig.~\ref{fig:frequ-fcc-hcp-fct-phi}}. ({In general, for} a random variable $X$ with a probability density $f(X)$, the symbol $[X]_p$ denotes the $p$-quantile, i.e.\ the value $F^{-1}(p)$ with the cumulative distribution function $F(X)=\int_{-\infty}^X f(\xi) {\rm d}\xi$.) The accuracy of our results is, for small $n_{\rm fcc}$ and $n_{\rm hcp}$, limited by the finite system size, preventing the measurement of probabilities smaller than $1/N$.  (We further note, that the number of isotropic cells, in the terminology of
Ref.~\cite{EPL10_SchroederTurk}, vanishes below $\phi_{\rm c}$, and becomes nonzero
at $\phi_{\rm c}$.)

\begin{figure}
\includegraphics[width=\columnwidth]{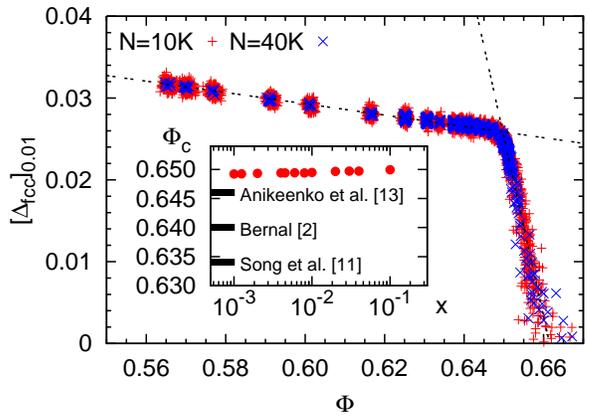}
\caption{\label{fig:fcc-percentile}(Color online) The fcc crystalline order metrics $\Delta_{\rm fcc}$ of the most fcc-like cells in each packing, quantified by the first percentile $[\Delta_{\mathrm{fcc}}]_{x}$ with $x=0.01$ of the $\Delta_{\rm fcc}$ distribution in each individual packing. (Each datapoint represents one packing). The insert shows variations of the estimate for $\phi_{\rm c}$  obtained as the intersection point of fitted straight lines, when $x$ is varied. The black bars indicate published estimates of $\phi_{\rm RCP}$: (a) Anikeenko and Medvedev's analysis of tetrahedral configurations \cite{PRL07_Anikeenko,AnikeenkoMedvedevAste:2008}, (b) Bernal's analysis of steel ball bearings \cite{Bernal:1959}, and (c) the contact number analysis by Song {\em et al.} \cite{nature08_Song}.}
\end{figure}

The values of $n_{\rm fcc}$ and $n_{\rm hcp}$ depend, of course, on the choice of the threshold $\delta$.
The increase in local crystallinity is, however, also evident in the lowest occurring values of $\Delta_{\mathrm{fcc}}$. Figure~\ref{fig:fcc-percentile} shows the first percentile $[ \Delta_{\mathrm{fcc}}]_{0.01}$ as a robust estimate for the most {crystal-like} cell, i.e.~the lowest occuring value of $\Delta_{\mathrm{fcc}}$. A sharp drop of $[ \Delta_{\mathrm{fcc}}]_{0.01}$ is observed for $\phi\gtrsim \phi_{\rm c}$; in packings below $\phi_{\rm c}$, the most fcc-like cells are substantially different from fcc cells, while above $\phi_{\rm c}$ the differences quickly decay to close to zero. The value of $\phi_{\rm c}(x)$ is estimated by the intersection of two straight lines fitted to the data for {$[\Delta_{\rm fcc}]_x$. The insert of Fig.~\ref{fig:fcc-percentile} shows the $\phi_{\rm c}$ estimates extracted by this approach, giving $\phi_{\rm c}\in [0.6492,0.6499]$ for $x\in [0.001,0.1]$. These values of $\phi_{\rm c}$ are} larger than published values for the RCP or the MRJ packing fraction.
Importantly, the data of Fig.~\ref{fig:fcc-percentile} demonstrate that the drastic increase in $n_{\mathrm{fcc}}$ in Fig.~\ref{fig:frequ-fcc-hcp-fct-phi} is a robust result that is not sensitive to the value of the threshold $\delta$. The value of $\delta$ is, within bounds,
an irrelevant parameter. We do not observe differences between packings of
$N=10^4$ and $N=4\times 10^4$ particles besides improved statistics.

The observed abrupt appearance of crystalline cells at $\phi_{\rm c}$ 
is difficult to observe using the bond-orientational order metrics $q_l$ and
$w_l$ developed in seminal work by Steinhardt {\em et al.}~\cite{PRB83_Steinhardt}.
Most frequently, $q_6$ is considered, which is deemed particularly sensitive to
formation of fcc; it is defined by
$$
     q_{6}(k)= \left[\frac{4\pi}{13}\sum_{m=-6}^{6}\left\vert\left< Y_{6m}\left(\theta_{jk},\varphi_{jk}\right) \right>\right\vert^2\right]^{1/2}
$$
with the spherical harmonics $Y_{lm}$, the polar angles $\theta_{jk}$ and $\varphi_{jk}$ of the bond vector between particles $j$ and $k$, and $\left< \cdot \right>$ denoting the average over the 12 closest neighbors $j$ of $k$ (Fig.~\ref{fig:neighbors-voronoi}).

\begin{figure}[t]
\centering
\begin{minipage}{1.0\columnwidth}
  \setlength{\unitlength}{\columnwidth}
  \begin{picture}(1,0.8333)
    \put(0.67,0.18){\includegraphics[width=0.24\columnwidth]{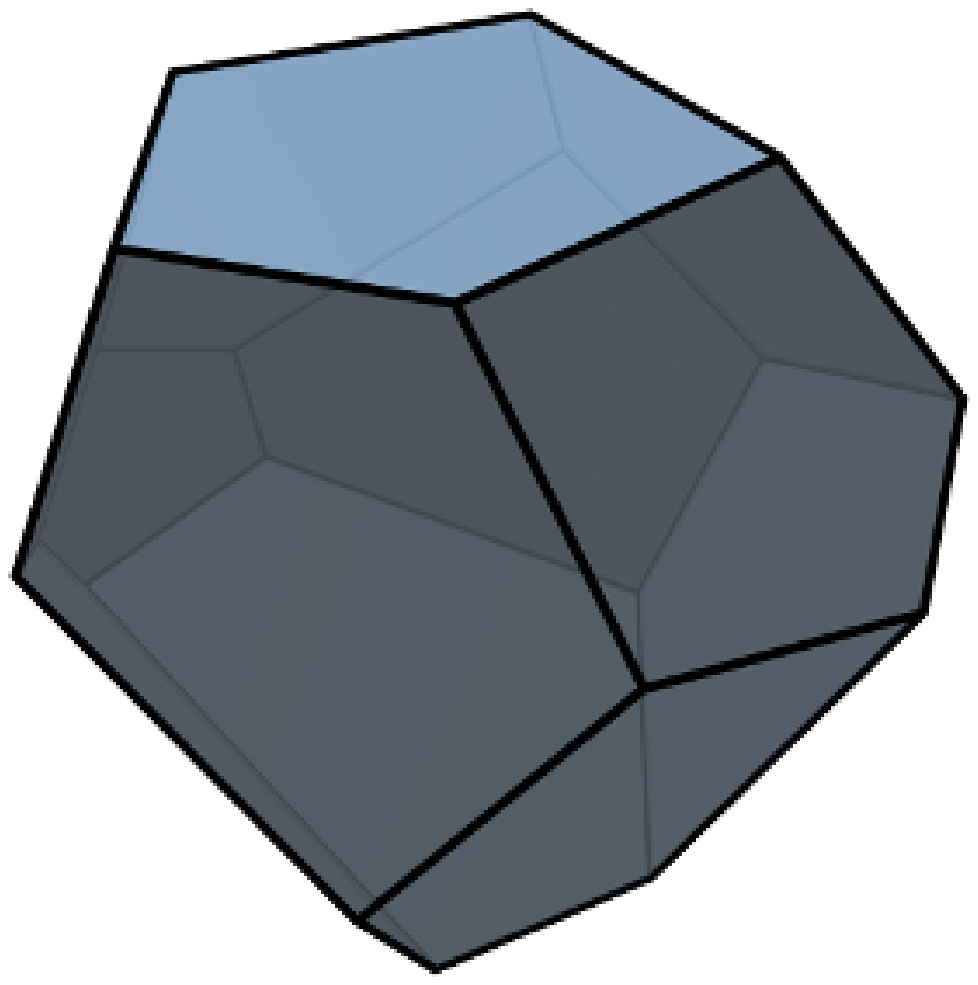} }
    \put(0.86,0.20){(d)} 
    \put(0.61,0.44){\includegraphics[width=0.32\columnwidth]{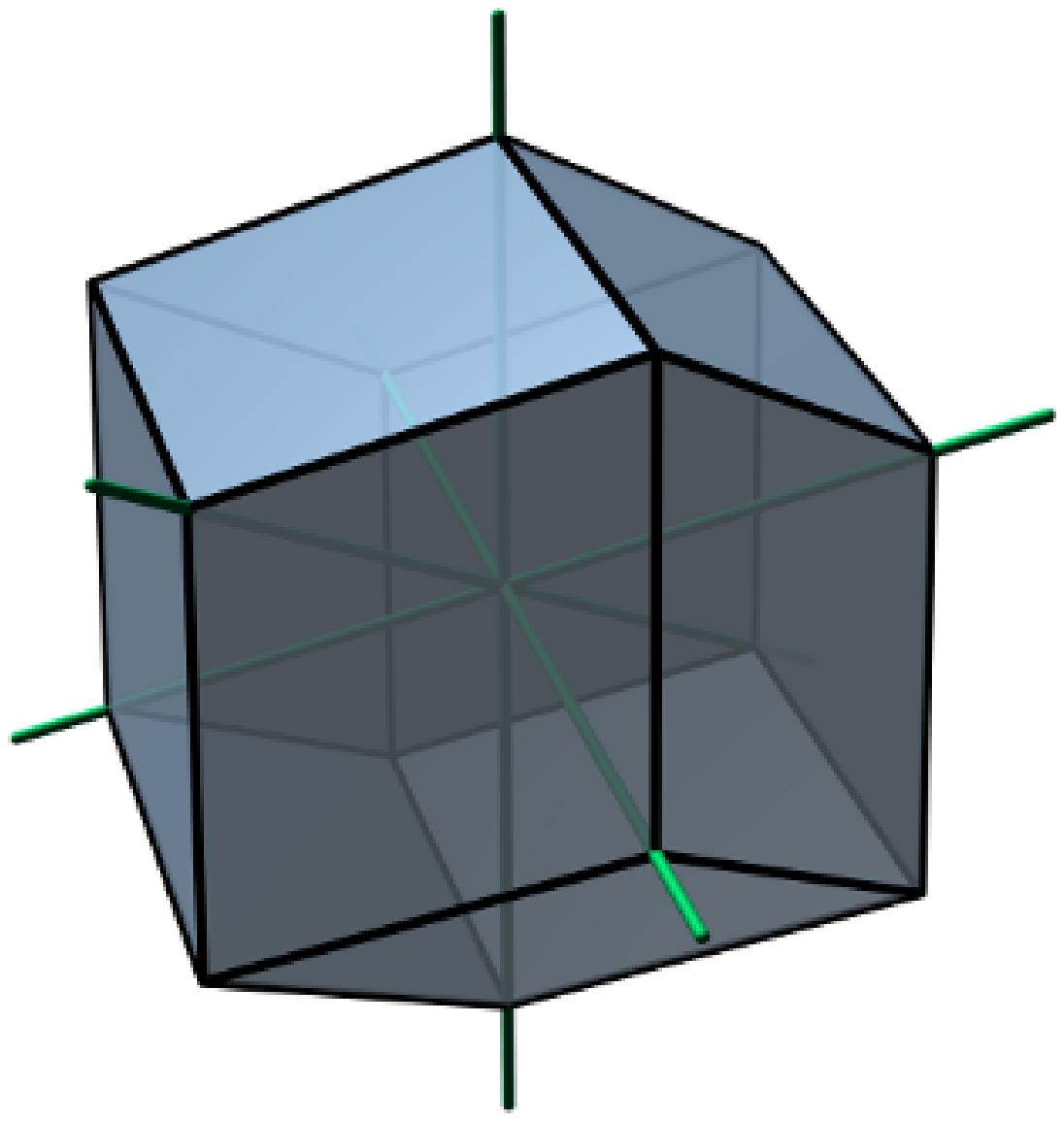} }
    \put(0.85,0.71){(b)}
    \put(0.12,0.45){\includegraphics[width=0.30\columnwidth]{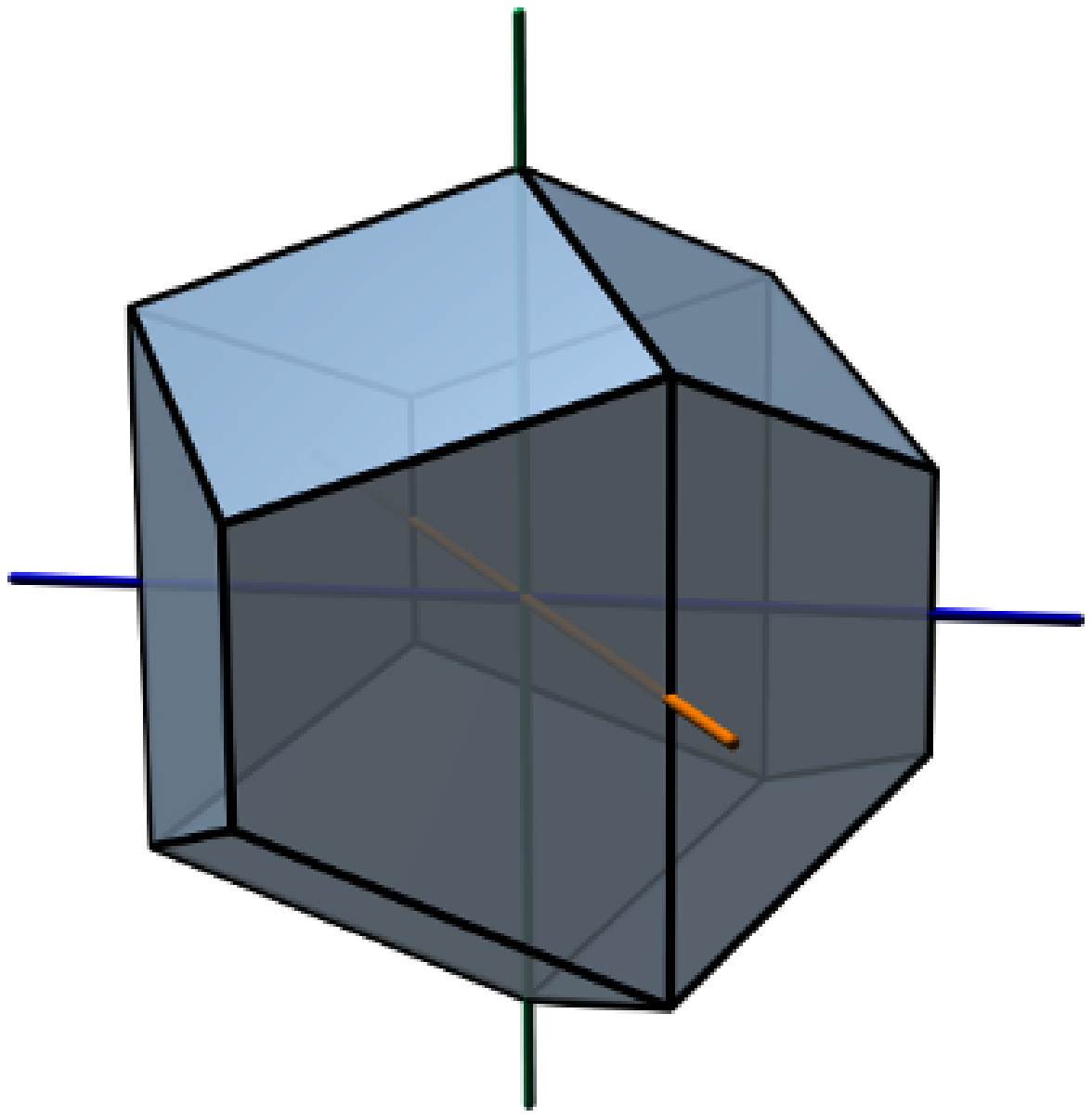} }
    \put(0.17,0.71){(a)}
    \put(0.255,0.17){\includegraphics[width=0.21\columnwidth]{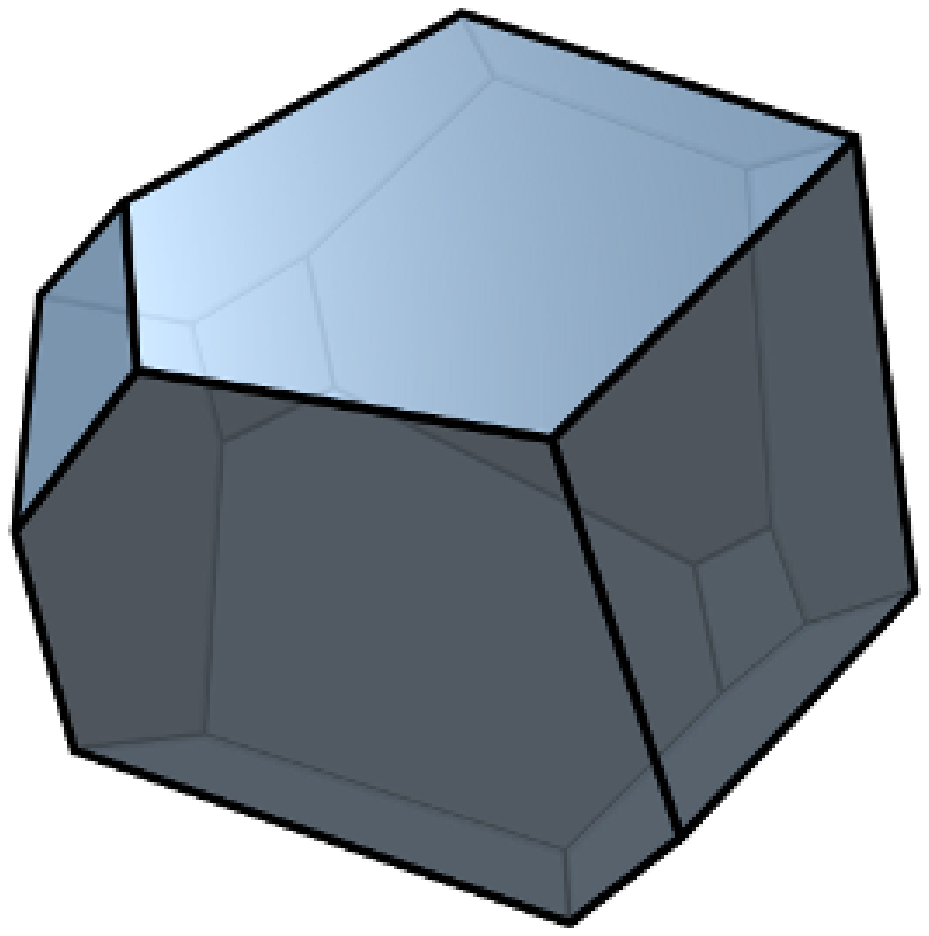} }
    \put(0.36,0.39){(c)}
    \put(0,0){\includegraphics[width=\columnwidth]{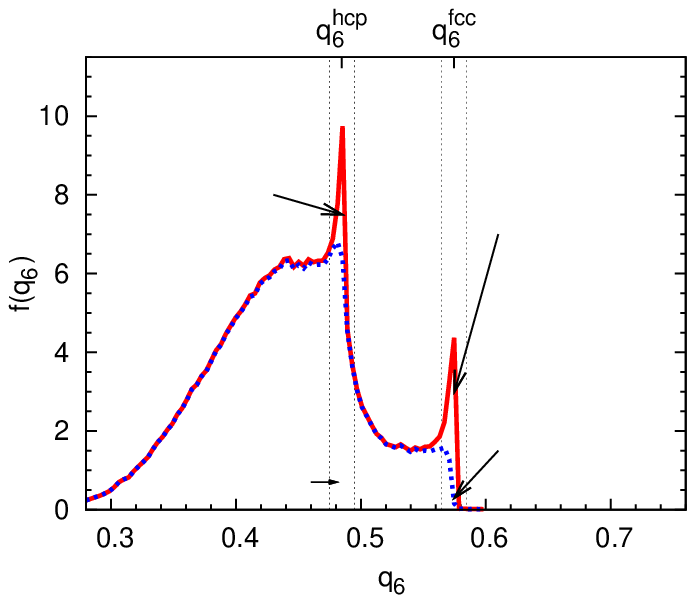}}
  \end{picture}
\end{minipage}
\caption{\label{fig:q6-distribution} (Color online) %
Abundances of {\em all} cells with a specific $q_6$ value ($f_{\rm all}$, red solid curve) and of the subset of
cells that are very clearly neither fcc nor hcp, i.e.~with $\Delta_{\rm fcc}>0.015$ and $\Delta_{\rm hcp}>0.015$ ($f_{\rm fp}$, blue dotted curve). The values of the blue, dotted curve at $q_6^{\rm fcc}$, $q_6^{\rm hcp}$ are finite even though genuine fcc and hcp cells are excluded. This clearly demonstrates that $q_6$ produces {\em false positives} (fp), that is cells that are not crystalline but identified as such by $q_6$.
Only the difference $f_{\rm cr} := f_{\rm all} - f_{\rm fp}$ consists of truly crystalline cells.  The cells depicted 
represent (a) an ideal hcp cell, (b) an ideal fcc cell, and cells identified by $q_6$, but not by $\Delta_{\rm fcc}$ or $\Delta_{\rm hcp}$, as (c) hcp and (d) fcc. The data is averaged over ten configurations, each consisting of $N=4\times 10^4$ spherical particles, with packing fractions $\phi$ in the interval $[0.656,0.660]$, well above $\phi_{\rm c}$. See Fig.~\ref{fig:neighbors-voronoi} for the definition of particle neighborhood.}
\end{figure}

Figure \ref{fig:q6-distribution} shows probability distributions of $q_6$
values observed in LS packings above $\phi_{\rm c}$ that demonstrate the
principal deficiency of using only $q_6$ as an order metric.
The frequency $f(q_6)$ of $q_6$ values develops sharp peaks
at the values corresponding to fcc ($q_6=0.57452$) and hcp ($q_6=0.48476$) for
$\phi \gtrsim \phi_{\rm c}$, not present for samples with $\phi <\phi_{\rm c}$.
These peaks, however, sit on top of a dominant background of non-crystalline cells.
The data clearly show that cells ({\em false positives}) exist which are distinctly different
from fcc but that are identified as fcc by $q_6$,
i.e.~$|q_6-q_6^{\rm fcc}|<5\cdot10^{-4}$. For example, the cell displayed in (d)
has eleven facets, several of which are five-sided; analogous hcp examples exist. 
If cells that are identified as either fcc or hcp by $W_1^{0,4}$
are excluded from the $q_6$ distribution, these peaks vanish;
the residual smooth distribution represents the non-crystalline background. 
Thus, for reliable detection of crystallinity, more information is required
than contained in $q_6$ alone.  This can be achieved 
by using multiple $q_l$ metrics and their distributional properties
\cite{KlumovKhrapakMorfill:2011,*KlumovUspekhi:2010}, or more specialized bond-orientational order
metrics such as $\theta^{\rm fcc}$, $\theta^{\rm hcp}$ \cite{Bargiel:2001:0921-8831:533}.
Minkowski tensors, such as used in the present study, represent a more general approach;
it is not necessary to choose a set of neighbors or bonds associated with a particle in
order to evaluate the Minkowski tensors, and the Minkowski tensors are continuous functions
of the particle coordinates. At the same time, they contain the information
necessary to discriminate in a robust and specific way between disordered structure
and different types of crystallinity.  It can be shown that a relation exists linking
the rotational invariants of higher-rank Minkowski tensors to variants of the bond-orientational
order metrics $q_l$, $w_l$ which have been amended by weighting factors proportional
to the Voronoi facet areas \cite{MickelKapferSchroeder-Turk:2011}.
    
In conclusion, we have demonstrated that {local crystallinity in Lubachevsky-Stillinger sphere packings sets in} when the packing is compactified beyond a {critical} packing fraction $\phi_{\rm c}\approx 0.649$. The packing fraction $\phi_{\rm c}$ marks the density below which LS configurations show no detectable degree of local crystallinity. Compactified above this limit, the system responds by the formation of local crystallinity.

The  value $\phi_{\rm c}\approx 0.649$ is higher than experimental estimates for the RCP limit \cite{BritJApplPhys69_Scott,Bernal:1959} and than the prediction based on mechanical contact numbers \cite{nature08_Song}, but also than the MRJ packing fraction \cite{PRL00_Torquato}. However, $\phi_{\rm c}$ is close to the packing fraction of $\approx 0.646$ where polytetrahedral aggregates are most prevalent (Fig.~3 in \cite{PRL07_Anikeenko}); the crystalline order metrics thus identify the conversion of polytetrahedral aggregates into crystalline structure, detected indirectly by Refs.~\cite{PRL07_Anikeenko,AnikeenkoMedvedevAste:2008}.
The small but significant discrepancy between the critical packing fractions $\leq 0.64$ of Ref.~\cite{BritJApplPhys69_Scott,Bernal:1959,nature08_Song,KamienLiu:2007} on the one hand and $\approx 0.65$ of Refs.~\cite{PRL07_Anikeenko,AnikeenkoMedvedevAste:2008} and of the present work on the other raise the caution that the mechanisms of dynamic arrest and isostaticity may be distinct from the alleged geometric order-disorder transition.

Given the nonequilibrium nature of jammed matter, one might be tempted to
attribute the observed difference in packing fraction to details of the preparation
protocol.  The critical packing fraction of $\approx 0.65$ is,
however, not specific to the LS algorithm.  For example,
for packings generated using the force-balancing 'split algorithm', the geometric (rather than
mechanical) contact number exhibits an anomaly close to $\phi\approx 0.65$
(Fig.~14 of Ref.~\cite{Jin:2010}). Data by Bargiel and Tory 
for Jodrey-Tory packings can be successfully fitted with a critical packing fraction of
$\approx 0.6495$ \cite{Bargiel:2001:0921-8831:533}.
Recent results by Klumov et al.\ for Jodrey-Tory and Lubachevsky-Stillinger packings,
in terms of quantiles of the $w_6$ distribution, fix the geometric transition around $\phi\approx 0.65$
\cite{KlumovKhrapakMorfill:2011,*KlumovUspekhi:2010}, but do not exclude crystallinity below $\phi_{\rm c}$.

Future work needs to focus on the precise nature of the geometric transition
occurring at $\phi_{\rm c}$.  Is the first-order phase transition scenario viable, and if so,
what are the coexistence densities?  What is the signature of the transition
in the $\Delta_{\rm fcc}$ distribution?  How does the local structure (quantified by Minkowski tensors)
relate to the observed ``Kauzmann'' density (Fig.~8 of Ref.~\cite{AnikeenkoMedvedevAste:2008})?
\vspace{.4cm}

\begin{acknowledgments}
We thank Matthias Schr\"oter, Tomaso Aste, Gary Delaney and Frank Rietz for insightful discussion, and the authors of Ref.~\cite{SkogeDonevStillingerTorquato:2006} for publishing their Lubachevsky-Stillinger implementation. We acknowledge support by the DFG through the research group ``Geometry \& Physics of Spatial Random Systems'' under grant SCHR 1148/3-1.  
\end{acknowledgments}

\bibliography{literature_prl}

\end{document}